\documentclass{article} 
\usepackage{iclr2026_conference,times}

\usepackage{amsmath,amsfonts,bm}

\def\eqref#1{equation~\ref{#1}}

\def\1{\bm{1}}

\def\ve{{\bm{e}}}

\def\vm{{\bm{m}}}

\def\vs{{\bm{s}}}

\def\mE{{\bm{E}}}

\def\mH{{\bm{H}}}

\DeclareMathAlphabet{\mathsfit}{\encodingdefault}{\sfdefault}{m}{sl}
\SetMathAlphabet{\mathsfit}{bold}{\encodingdefault}{\sfdefault}{bx}{n}

\usepackage{hyperref}
\usepackage{url}
\usepackage{graphicx}
\usepackage{booktabs}
\usepackage{multirow}
\usepackage[table]{xcolor}
\usepackage{amssymb}
\usepackage{colortbl}
\usepackage{caption}
\usepackage{wrapfig}  
\usepackage{booktabs} 
\usepackage{diagbox}

\usepackage{graphicx}
\usepackage{subcaption}

\title{FakeMark: Deepfake Speech Attribution With Watermarked Artifacts}

\author{Wanying Ge, Xin Wang, Junichi Yamagishi \\
National Institute of Informatics\\
Chiyoda-ku, Tokyo 101-8430, Japan\\
\texttt{\{gewanying,wangxin,jyamagis\}@nii.ac.jp}
}

\iclrfinalcopy 
\begin{document}

\maketitle

\begin{abstract}
Deepfake speech attribution remains challenging for existing solutions. Classifier-based solutions often fail to generalize to domain-shifted samples, and watermarking-based solutions are easily compromised by distortions like codec compression or malicious removal attacks. To address these issues, we propose FakeMark, a novel watermarking framework that injects artifact-correlated watermarks associated with deepfake systems rather than pre-assigned bitstring messages. This design allows a detector to attribute the source system by leveraging both injected watermark and intrinsic deepfake artifacts, remaining effective even if one of these cues is elusive or removed. Experimental results show that FakeMark improves generalization to cross-dataset samples where classifier-based solutions struggle and maintains high accuracy under various distortions where conventional watermarking-based solutions fail. Speech samples are available at~\url{https://fakemark-demo.github.io/fakemark-demo/}.

\end{abstract}

\section{Introduction}
\label{sec:intro}

\begin{figure}[t]
    \centering
    \includegraphics[width=1\linewidth, trim=10 0 0 0, clip]{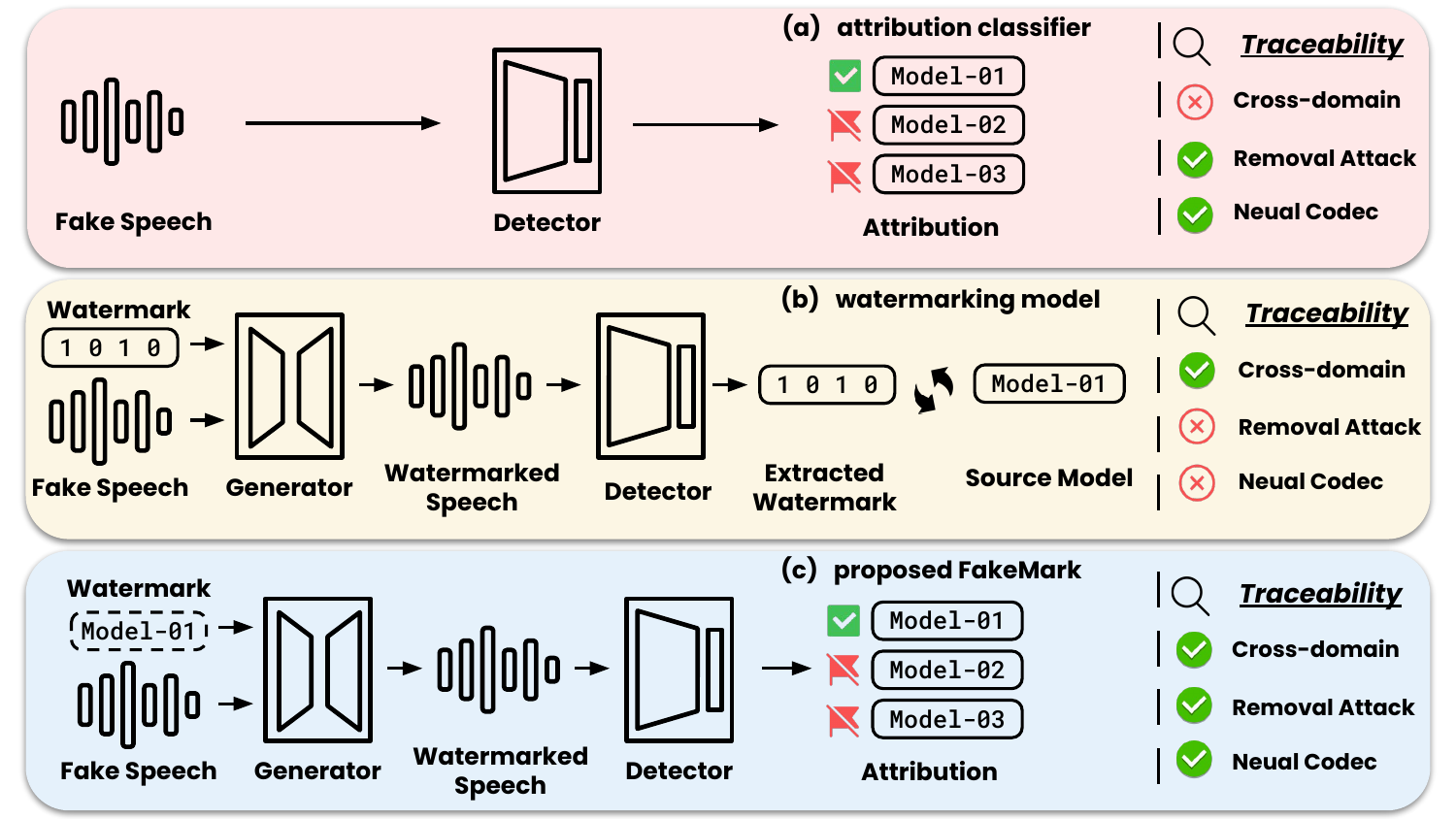}
    \caption{Illustrations of deepfake attribution by (a) classifier-based model, (b) watermarking-based model, and (c) our proposed FakeMark.}
    \label{fig:comparsion}
\end{figure}

Attributing deepfake speech requires identifying the source system used to generate the synthetic samples~\citep{muller2022attacker, klein2025open}. This is critical for mitigating risks such as copyright violations and malicious use of speech synthesis systems. Most solutions train deep-neural-network-based classifiers (illustrated in the top panel of Figure~\ref{fig:comparsion}) for system identification~\citep{sun2023vocoder,wang2025generalize}. They often require to be trained in a discriminative manner with data generated by a rich variety of speech synthesis systems to capture artifact differences. However, such classifiers are known to be sensitive to domain shift and struggle to detect deepfakes generated by unseen systems~\citep{bhagtani2024attribution, chen25codecsc}, as their performance is fundamentally constrained by the variability present in the training data.

Recently, watermarking-based methods become popular as an alternative solution to the attribution task~\citep{cho2022attributable,li2025trinimark,yang2025dualmark}. These solutions involve training a pair of watermark generator and detector (illustrated in the middle panel of Figure~\ref{fig:comparsion}), where the generator injects a watermark message into the carrier speech that is later extracted by the detector; attribution is achieved by mapping the extracted message to its pre-assigned system label. Although watermarking-based solutions have demonstrated high accuracy on various benchmarks~\citep{liu2024audiomarkbench,sanroman2024proactive}, they can be easily compromised by common distortions and removal attacks~\citep{yang2024can,kassis2025unmarker,yao2025yours}. In its application to speech, for example, watermark generators are trained to inject watermarks that are inaudible to the human ear. Yet watermark detectors often struggle under neural codec transmissions~\citep{juvela2025audio, oreilly2025deep, ozer25survive}, whose training objective is compression and high-fidelity reconstruction of audio signals~\citep{defossez2022encodec,ju2024naturalspeech}. In deepfake related tasks such as detection~\citep{wu25comparative}, classifier-based solutions remain robust under neural codecs since deepfake artifacts are preserved to some extent, whereas watermark detectors degrade to near-chance performance as the injected messages are removed during compression.

Presented in this paper is our attempt to address the above challenges for robust deepfake speech attribution. Specifically, we ask the following research question: \textbf{Can we enhance deepfake traceability by injecting artifact-correlated watermarks}? We hypothesize that correlating watermarks with deepfake artifacts can provide improved 1) \textbf{robustness towards distortions} by enabling the detector to perform attribution through acoustic artifacts when the watermark message is removed; and 2) \textbf{generalization performance} by introducing a watermark generator to assist the typically standalone classifier-based detector when seen artifact patterns are absent. To answer the question, we propose a novel watermarking framework (illustrated in the bottom panel of Figure~\ref{fig:comparsion}) and evaluate its performance under diverse conditions. Our main contributions are summarized as follows:
\begin{itemize}
    \item We introduce FakeMark, a novel watermarking framework for deepfake speech attribution. The FakeMark generator injects watermarks that are correlated with acoustic artifacts, allowing the detector to map either the artifacts or the watermarks to their source system.
    \item We present the first systematic evaluation of deepfake speech attribution using both watermarking- and classifier-based models. We evaluate FakeMark against these baselines on common datasets and under diverse distortions, showing that it improves attribution robustness and generalization in challenging scenarios.
\end{itemize}

\section{Related works}

\textbf{Speech generation} typically follows two paths: text-to-speech (TTS) and voice conversion (VC). In modern TTS, an acoustic model maps the input text (or its derived linguistic features) to an intermediate acoustic representation that is either continuous valued hidden feature vectors or discrete tokens. A neural vocoder is then used to synthesize the speech waveform~\citep{tan2021survey}. VC follows a partially similar design: it takes an input waveform from a source speaker and renders the same content in a target speaker’s voice. The term \emph{artifacts} denotes deviations of synthesized speech from natural speech. Common audible artifacts include (i) alignment errors between text and predicted acoustics that cause word skipping or repetition~\citep{zen2009statistical}; (ii) insufficient modeling of prosody (e.g., incorrect pitch accent~\citep{fastpitch}), expressiveness (e.g., flat intonation~\citep{liu2021expressive,mahapatra25can})
, and speaker characteristics (e.g., a voice perceptually dissimilar to the target speaker~\citep{chen2025valle,pan22speaker}); and (iii) vocoder artifacts such as buzziness or high-frequency noise~\citep{bak2023avocodo,sun2023vocoder}.

\textbf{Deepfake attribution.} Depending on the specific architectures used, it has been reported that different acoustic models~\citep{bhagtani2024attribution,chen2025towards} and vocoders~\citep{sun2023vocoder,deng2024vocoderartifacts} leave distinctive artifacts that can be leveraged for deepfake attribution. Solutions to the task naturally involve collecting samples from various TTS and VC systems~\citep{muller2024mlaad,chen25codecsc}. These samples are then used to train multiclass classifiers to predict the acoustic models or vocoders used for their generation~\citep{klein24source}. However, such supervised training scheme can sometimes cause classifiers to exploit undesired differences in the training data, leading to poor generalization performance on unseen samples. This includes samples generated by seen systems but trained with different languages~\citep{marek2024audio} or speakers~\citep{klein2025open}, or samples with subtle artifacts like unnatural silences~\citep{chen2025towards} or even generated by the same system with different weights~\citep{stan25tada}. Alternative strategies to achieve generalization include estimating model confidence on unseen samples~\citep{klein2025open} or measuring sample similarities in latent space, akin to verification tasks~\citep{negroni25sv}.

\textbf{Speech watermarking} models are designed to inject and extract bitstring messages within a speech signal~\citep{li2025trinimark,liu2024audiomarkbench}. Depending on the information carried in the message, these models are versatile for various tasks. For example, a watermark can encode a compressed version of the original signal for self-recovery~\citep{quinonez2024_selfrecovery}, or it can be assigned to different users for attribution and copyright protection~\citep{sanroman2024proactive,timbrewatermarking}. In deepfake-related applications, different bitstrings can be assigned to real and fake samples for detection~\citep{wu25comparative,sanroman2024proactive} or to counter malicious deepfake manipulations~\citep{li25voicemark,he2025assmark}. Beyond bitstring messages, the presence of a watermark itself can represent a zero-bit message indicating whether a sample is real or fake~\citep{juvela2024collaborative,roman2025latent}. Previous studies have reported that watermarking-based models are vulnerable to distortions such as neural codecs, malicious forgery, or removal attacks~\citep{yang2024can,liu2024audiomarkbench}. 
Common strategies to enhance robustness include using codec-based data augmentation during training~\citep{juvela2025audio} and injecting watermarks into deep latent representations of the speech signal~\citep{ji2024speechwatermark}.

\section{FakeMark}

We describe FakeMark pipelines for watermark generation and detection in Sec.~\ref{sec:watermark_generation}. Objectives used to train the system modules are detailed in Sec.~\ref{sec:watermark_detection}.

\subsection{Pipeline}
\label{sec:watermark_generation}

\begin{figure}
    \centering
    \includegraphics[width=1\linewidth,trim=0 50 20 20, clip]{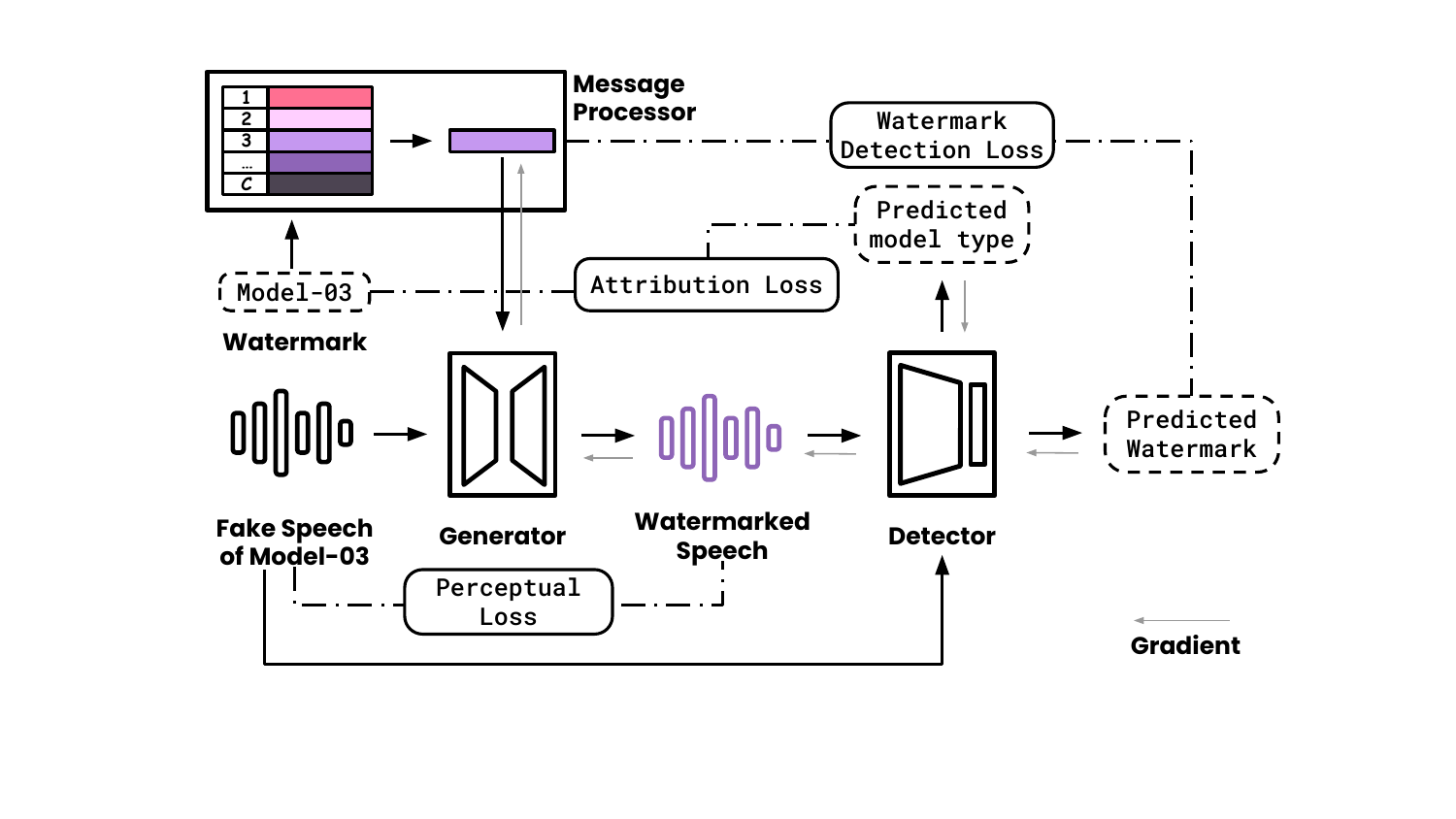}
    \caption{Training pipeline of FakeMark.}
    \label{fig:training}
\end{figure} 

As illustrated in Figure~\ref{fig:training}, FakeMark takes two inputs during watermark generation: the speech signal $\vs \in \mathbb{R}^{T}$ and the watermark message $w \in \{1,\dots,C\}$, where $T$ is the number of waveform sampling points and $C$ is the total number of deepfake systems. It outputs a watermarked signal $\vs_{w} \in \mathbb{R}^{T}$ that carry the watermark message and has the same dimensionality as the input signal. During watermark detection, FakeMark takes $\vs_{w}$ and predicts a watermark message $w'$.

The generation process involves four stages:

\begin{enumerate}
    \item Given the input watermark message $w$, the message processor returns the corresponding embedding vector $\ve_{w} \in \mathbb{R}^{H}$ to the generator, where $H$ is the latent dimension.
    \item Given the input signal $\vs$, the generator down-samples it and extracts a compact latent representation $\mH_{s} \in \mathbb{R}^{\lfloor \tfrac{T}{\alpha} \rfloor \times H}$, where $\alpha$ is the downsampling factor.
    \item Given the input watermark embedding $\ve_{w}$, the generator repeats it along the time axis to form $\mE_{w} \in \mathbb{R}^{\lfloor \tfrac{T}{\alpha} \rfloor \times H}$, then applies voice-activity detection (VAD) to obtain a binary mask $\vm \in \{0,1\}^{\lfloor \tfrac{T}{\alpha} \rfloor}$ indicating speech-active frames, and computes the watermark latent $\mH_{w} = \vm \odot \mE_{w}$, where $\mH_w \in \mathbb{R}^{\lfloor \tfrac{T}{\alpha} \rfloor \times H} $.
    \item The generator up-samples $\mH_{s}+\mH_{w}$ and outputs the watermark signal $\boldsymbol{\delta}_{w} \in \mathbb{R}^{T}$. The final watermarked signal is obtained as $\vs_{w} = \vs+\boldsymbol{\delta}_{w}$, where $\vs_{w} \in \mathbb{R}^{T}$.
\end{enumerate}

Following the generation pipeline, we explore two generator architectures:
\begin{itemize}
    \item FakeMark$^{A}$: follows an encoder-decoder architecture that processes speech waveforms, as used in AudioSeal~\citep{sanroman2024proactive};
    \item FakeMark$^{T}$: follows an encoder architecture that processes spectrogram features, as used in Timbre~\citep{timbrewatermarking}. In this setting, $\mH_{s}$ is the linear-scale spectrogram obtained via Short-Time Fourier Transform (STFT). Final waveforms are obtained via inverse STFT.
\end{itemize}

The detection process involves two stages:
\begin{enumerate}
    \item During training, given the input watermarked signal $\vs_{w}$, the detector applies a series of transformations to obtain a distorted version input $\vs'_{w}$. This strategy ensures robustness of the watermark injection and detection. Full list of the used transformations and their settings are provided in Appendix~\ref{apd:augmentation}. These transformations are disabled during inference.
    \item The detector extracts sequence-level feature from the input waveform and predicts the class probabilities $\mathbf{p}\in[0,1]^C$ over $C$ watermark types; the extracted watermark is obtained as $w'=\arg\max_{i\in\{1,\dots,C\}} p_i$.
\end{enumerate}

We use a common detector architecture that consists of a pre-trained SSL front-end~\citep{pratap2024_mms} and a fully connected back-end classifier. Detailed architectures are provided in the Appendix~\ref{apd:architecture}.

\subsection{Training objectives}
\label{sec:watermark_detection}

All FakeMark modules are optimized with three classes of objectives: 1) attribution loss, to ensure the ability of distinguishing different types of artifacts; 2) detection loss, to maximize the successful injection and detection of watermarks; and 3) perceptual loss, to minimize the perceptual distortion between original and watermarked signals. They are detailed below.

\textbf{Attribution loss} differentiates FakeMark from conventional watermarking approaches. It is computed as the cross-entropy between the ground-truth deepfake system label and the detector's predicted class probabilities over an \emph{unwatermarked} clean signal. This objective is similar to training classifier-based attribution models~\citep{klein24source,sun2023vocoder}, where the goal is to capture distinct characteristics of deepfakes generated by different systems. The back-propagated attribution loss encourages the detector to distinguish various types of deepfake artifacts and implicitly guides the watermark embeddings from the message processor to correlate with these artifacts. As a result, each watermark embedding encodes the artifact patterns learned from all samples of a specific deepfake system in the training set.

\textbf{Watermark detection loss} ensures that the generator injects watermark messages that can be reliably recognized by the detector. Similar to the training of conventional watermarking models, a \emph{random} watermark message is sampled and processed to obtain a watermark embedding, which is then used for watermarked signal generation. The detector predicts class probabilities from this \emph{watermarked} signal, and the watermark detection loss is computed as the cross-entropy between the ground-truth watermark message and the predicted distribution.

By jointly training FakeMark with both attribution and watermark detection losses, the message processor learns to align watermark embeddings with the deepfake artifacts they represent, and the generator-detector pair learns to robustly inject and detect these watermarks. During inference, with the watermark message always chosen to match the ground-truth deepfake system, FakeMark can attribute the source system using both acoustic artifacts and the watermarks. This ensures effective attribution even if one of these cues is compromised.

\textbf{Perceptual losses} promote the imperceptibility of watermarks and the naturalness of watermarked signals. This is achieved by refining the watermarked signal with HiFi-GAN–style losses~\citep{kong2020hifi}, which include a Mel-spectrogram reconstruction loss to enforce the spectral similarity and adversarial discriminator losses to improve speech fidelity.

Additionally, we follow~\cite{sanroman2024proactive} to refine the watermark signals generated by both generator architectures. We apply $l_1$ loss and loudness on the watermark signal to decrease its intensity and ensure its robustness towards distortions. We further use a frequency magnitude loss to align the averaged spectral envelope of the watermark signal with that of the clean signal, promoting perceptual similarity and ensuring the watermarks remain less audible.

\section{Experiments and Results}

We perform in-domain evaluation with seen artifacts and cross-dataset evaluation with unseen artifacts. FakeMark is compared against recent baselines on both clean and distorted signals. In addition to attribution accuracy, we also assess the speech quality of watermarked signals.

\subsection{Experimental setup}

\textbf{Datasets.} We use the MLAAD\_v5 dataset~\citep{muller2024mlaad} for training and evaluation. Following the source tracing challenge protocol~\citep{UsingMLAADforSourceTracing}, our training set comprises 24 TTS systems covering eight languages. To mitigate the influence of undesirable artifacts related to language or speaker~\citep{klein24source}, we group systems with identical architectures into a single class. The resulting training set contains 12 classes, 9 of which appear in the evaluation set. For cross-dataset evaluation, we collected samples generated by five of these systems from ASVspoof5~\citep{wang2024_asvspoof5} and TIMIT-TTS~\citep{salvi2023timittts} datasets. Both evaluation sets were randomly sampled with an equal number of files per class. All evaluated system architectures are seen during training; evaluations on unseen architectures are beyond the scope of this work. Dataset details are provided in Appendix~\ref{apd:dataset}.

\textbf{Baseline systems.} We compare our FakeMark with recent and remarkable watermarking models: AudioSeal~\citep{sanroman2024proactive} and Timbre~\citep{timbrewatermarking}, and a ResNet34-based classifier~\citep{klein2025open} that takes STFT spectrograms as input. Additionally, we train an SSL-based classifier with the same architecture as FakeMark detector (denoted MMS-300M) to isolate the impact of the watermarking scheme. All baselines were trained with the same training set and augmentation strategy as FakeMark. Watermark message length for AudioSeal and Timbre is set to 4 (equivalent to $2^4$ unique bitstrings)-the minimum capacity required to cover 12 classes. Full model configurations and implementation details are in Appendix~\ref{apd:training}.

\textbf{Evaluation metrics.} Objective evaluation for both attribution and audio quality are performed to evaluate FakeMark and baselines:
\begin{itemize}
    \item For attribution performance, we report accuracy result. Predicted class for AudioSeal and Timbre is the class whose assigned 4-bit message has the shortest Hamming distance to the detector output~\citep{sanroman2024proactive, liu2025xattn}. For FakeMark and classifier-based models, predicted class is the class with the highest detector probability.
    \item For audio quality assignment of watermarked signals, we use four different metrics: Scale Invariant Signal to Noise Ratio (SI-SNR) for evaluating noise-level of watermarks; PESQ to evaluate speech quality for telecom-like scenarios~\citep{pesq}; ViSQOL for assessing perceptual quality for network-based scenarios~\citep{visqol}; Production Quality (PQ) to estimate the clarity and fidelity of watermarked signals~\citep{tjandra2025meta}. Unlike the previous three metrics, PQ does not require clean reference signals.    
\end{itemize}

\textbf{Distortions and attacks.} To evaluate system robustness against distortions and watermark removal attacks, we apply a set of transformations previously shown to have a noticeable impact on either watermark extraction~\citep{oreilly2025deep,yao2025yours,yang2024can} or deepfake detection~\citep{wu25comparative} in the literature. They are applied to the watermarked signals for FakeMark, AudioSeal, and Timbre, and directly to the input signals for ResNet34 and MMS-300M, include:
\begin{itemize}
    \item Signal processing-based transforms: Pitch shift, playback speed change, and additive noise from MUSAN~\citep{snyder2015_musan};
    \item Neural Codec-based waveform compression and regeneration: SpeechTokenizer~\citep{zhang2023speechtokenizer}, FACodec~\citep{ju2024naturalspeech,zhang2023amphion}, and WavToenizer~\citep{ji2024wavtokenizer};
    \item Neural Vocoder-based waveform regeneration: HiFi-GAN~\citep{kong2020hifi}, Vocos~\citep{siuzdak2023vocos}, and BigVGAN~\citep{lee2022bigvgan}.
    \item Black-box watermark removal attacks: Overwriting~\citep{yao2025yours}, where publicly-available, pre-trained Timbre and AudioSeal models are run sequentially to overwrite existing watermarks; and Averaging~\citep{yang2024can}, where an average watermark is computed using a pre-trained AudioSeal model and then subtracted from the watermarked signals. 
\end{itemize}

    Details of distortions and attacks can be found in Appendix~\ref{apd:distortion}.

\subsection{Results}

We report deepfake attribution performance in this section, including in-domain evaluation results in Sec.~\ref{sec:in-domain} and cross-dataset evaluation results in Sec.~\ref{sec:out-domain}. Speech quality evaluation and additional analysis are reported in Sec.~\ref{sec:subj} and Sec.~\ref{sec:abla}.

\subsubsection{Evaluation with seen artifacts}
\label{sec:in-domain}

Table~\ref{tab:id} presents attribution accuracy for FakeMark and baselines on the MLAAD\_v5 test set. Rows represent accuracies under different distortions. Cells are color-coded in grayscale by row: darker shades indicating lower accuracy and lighter shades indicating higher accuracy. We summarize observations related to our research question below.

\begin{table*}[h]
\centering
\renewcommand{\arraystretch}{1.1}
\caption{
Attribution accuracy results on seen artifacts across distortions and attacks. 
}
\resizebox{0.99\textwidth}{!}{%
\begin{tabular}{llcccccc}
\toprule
\multirow{3}{*}{\textbf{}} &  \multirow{2}{*}{\diagbox{Distortion}{System}}
& \multicolumn{2}{c}{Proposed Method} 
& \multicolumn{2}{c}{Watermarking Baselines} 
& \multicolumn{2}{c}{Classifier Baselines} \\
\cmidrule(lr){3-4} \cmidrule(lr){5-6} \cmidrule(lr){7-8}
& & \textbf{FakeMark$^{A}$} & \textbf{FakeMark$^{T}$} & \textbf{AudioSeal} & \textbf{Timbre} 
& \textbf{MMS-300M} & \textbf{ResNet34} \\
\midrule
 & None             & \cellcolor[rgb]{1.00, 1.00, 1.00} 1.00 & \cellcolor[rgb]{1.00, 1.00, 1.00} 1.00 & \cellcolor[rgb]{1.00, 1.00, 1.00} 1.00 & \cellcolor[rgb]{1.00, 1.00, 1.00} 1.00 & \cellcolor[rgb]{1.00, 1.00, 1.00} 1.00 & \cellcolor[rgb]{0.59, 0.59, 0.59} 0.97\\
\midrule
\multirow{3}{*}{Signal Processing} 
& Pitch      & \cellcolor[rgb]{0.94, 0.94, 0.94} 0.82 & \cellcolor[rgb]{1.00, 1.00, 1.00} 1.00 & \cellcolor[rgb]{0.93, 0.93, 0.93} 0.80 & \cellcolor[rgb]{0.99, 0.99, 0.99} 0.96 & \cellcolor[rgb]{0.59, 0.59, 0.59} 0.27 & \cellcolor[rgb]{0.96, 0.96, 0.96} 0.88\\
& Speed      & \cellcolor[rgb]{1.00, 1.00, 1.00} 1.00 & \cellcolor[rgb]{1.00, 1.00, 1.00} 1.00 & \cellcolor[rgb]{0.59, 0.59, 0.59} 0.85 & \cellcolor[rgb]{0.95, 0.95, 0.95} 0.97 & \cellcolor[rgb]{1.00, 1.00, 1.00} 1.00 & \cellcolor[rgb]{0.84, 0.84, 0.84} 0.92\\
& Noise      & \cellcolor[rgb]{0.82, 0.82, 0.82} 0.63 & \cellcolor[rgb]{0.92, 0.92, 0.92} 0.71 & \cellcolor[rgb]{0.94, 0.94, 0.94} 0.72 & \cellcolor[rgb]{0.78, 0.78, 0.78} 0.60 & \cellcolor[rgb]{1.00, 1.00, 1.00} 0.80 & \cellcolor[rgb]{0.59, 0.59, 0.59} 0.50\\
\midrule
\multirow{3}{*}{Codec} 
& SpeechTokenizer & \cellcolor[rgb]{0.96, 0.96, 0.96} 0.85 & \cellcolor[rgb]{1.00, 1.00, 1.00} 0.99 & \cellcolor[rgb]{0.59, 0.59, 0.59} 0.10 & \cellcolor[rgb]{0.99, 0.99, 0.99} 0.94 & \cellcolor[rgb]{0.98, 0.98, 0.98} 0.92 & \cellcolor[rgb]{0.97, 0.97, 0.97} 0.88\\ 
& FACodec    & \cellcolor[rgb]{0.98, 0.98, 0.98} 0.91 & \cellcolor[rgb]{1.00, 1.00, 1.00} 0.99 & \cellcolor[rgb]{0.59, 0.59, 0.59} 0.17 & \cellcolor[rgb]{0.95, 0.95, 0.95} 0.82 & \cellcolor[rgb]{0.98, 0.98, 0.98} 0.92 & \cellcolor[rgb]{0.94, 0.94, 0.94} 0.79\\  
& WavTokenizer & \cellcolor[rgb]{0.80, 0.80, 0.80} 0.33 & \cellcolor[rgb]{0.81, 0.81, 0.81} 0.34 & \cellcolor[rgb]{0.59, 0.59, 0.59} 0.09 & \cellcolor[rgb]{0.69, 0.69, 0.69} 0.19 & \cellcolor[rgb]{0.84, 0.84, 0.84} 0.39 & \cellcolor[rgb]{1.00, 1.00, 1.00} 0.71\\  
\midrule
\multirow{3}{*}{Vocoder} 
& HiFi-GAN   & \cellcolor[rgb]{0.98, 0.98, 0.98} 0.91 & \cellcolor[rgb]{1.00, 1.00, 1.00} 1.00 & \cellcolor[rgb]{0.59, 0.59, 0.59} 0.09 & \cellcolor[rgb]{1.00, 1.00, 1.00} 1.00 & \cellcolor[rgb]{0.99, 0.99, 0.99} 0.94 & \cellcolor[rgb]{0.98, 0.98, 0.98} 0.92\\
& Vocos      & \cellcolor[rgb]{1.00, 1.00, 1.00} 0.98 & \cellcolor[rgb]{1.00, 1.00, 1.00} 1.00 & \cellcolor[rgb]{0.59, 0.59, 0.59} 0.12 & \cellcolor[rgb]{1.00, 1.00, 1.00} 1.00 & \cellcolor[rgb]{1.00, 1.00, 1.00} 0.98 & \cellcolor[rgb]{0.99, 0.99, 0.99} 0.97\\
& BigVGAN    & \cellcolor[rgb]{1.00, 1.00, 1.00} 0.99 & \cellcolor[rgb]{1.00, 1.00, 1.00} 1.00 & \cellcolor[rgb]{0.59, 0.59, 0.59} 0.28 & \cellcolor[rgb]{1.00, 1.00, 1.00} 1.00 & \cellcolor[rgb]{1.00, 1.00, 1.00} 1.00 & \cellcolor[rgb]{0.99, 0.99, 0.99} 0.97\\ 
\midrule
\multirow{2}{*}{Removal Attack}
& Overwriting & \cellcolor[rgb]{1.00, 1.00, 1.00} 0.99 & \cellcolor[rgb]{0.98, 0.98, 0.98} 0.95 & \cellcolor[rgb]{0.76, 0.76, 0.76} 0.68 & \cellcolor[rgb]{0.59, 0.59, 0.59} 0.55 & \cellcolor[rgb]{0.98, 0.98, 0.98} 0.95 & \cellcolor[rgb]{0.83, 0.83, 0.83} 0.75\\
& Averaging   & \cellcolor[rgb]{0.98, 0.98, 0.98} 0.98 & \cellcolor[rgb]{0.99, 0.99, 0.99} 0.99 & \cellcolor[rgb]{0.59, 0.59, 0.59} 0.79 & \cellcolor[rgb]{1.00, 1.00, 1.00} 1.00 & \cellcolor[rgb]{1.00, 1.00, 1.00} 1.00 & \cellcolor[rgb]{0.96, 0.96, 0.96} 0.96\\
\bottomrule
\end{tabular}
}
\label{tab:id}
\end{table*}

\textbf{FakeMark is robust to strong watermark removal distortions}. When no distortion is applied, all models achieve near-perfect accuracy (above 0.97). Across most distortions—except background noise and WavTokenizer—both FakeMark variants maintain high attribution accuracy (above 0.80). In contrast, AudioSeal accuracy drops dramatically under codec (0.09–0.17) and vocoder (0.09–0.28) reconstructions, which is expected given that these distortions are known strong watermark removers~\citep{oreilly2025deep, juvela2025audio}. Although FakeMark$^{A}$ shares the same generator architecture as AudioSeal, its detector can still leverage deepfake artifacts for attribution, yielding performance that is similar to that of the MMS-300M classifier.

Our baseline Timbre model demonstrates unexpectedly robust performance across distortions previously reported as vulnerabilities~\citep{oreilly2025deep,ozer25survive}. This is likely due to retraining with additional augmentations, including a codec method named EnCodec~\citep{defossez2022encodec}.

\textbf{FakeMark is robust to watermark removal attacks.} Though Timbre performance was enhanced with additional augmentations, both watermarking baselines remain vulnerable to removal attacks: Timbre drops from 1.00 to 0.55 under overwriting, and AudioSeal drops to 0.79 under averaging. Both FakeMark variants are less affected, with the lowest accuracy being 0.95—substantially more robust than the other watermarking baselines.

\textbf{Additional discussion: models process spectrogram features are generally more robust.} Table~\ref{tab:id} shows that attribution is easily solved under clean conditions. Even with distortions, most models—except AudioSeal—maintain reliable performance in many scenarios. We also notice that models process spectrogram features are more robust to distortions compared to their counterparts. Watermarking models such as FakeMark$^{T}$ and Timbre achieve perfect accuracies (1.0) under neural vocoders. The ResNet34 is the only solution that does not reach perfect performance under clean conditions (0.97); however, its lowest accuracy (0.50 under Noise) remains noticeably higher than the MMS-300M’s lowest results (0.27 under Pitch shift and 0.39 under WavTokenizer).

Almost all tested models appear sensitive to signal processing–based distortions but relatively more robust to other types of distortions. This is expected, as signal processing transforms directly modify the speech signal and thus alter artifact patterns. By contrast, reconstruction-based distortions primarily regenerate the waveform together with artifacts and, in some cases, watermarks. In the next section, we show that attribution becomes more difficult when the artifact patterns are unseen, particularly for the two classifier-based baselines.


\subsubsection{Evaluation with unseen artifacts}
\label{sec:out-domain}

Table~\ref{tab:od} present cross-dataset evaluation of attribution accuracy of FakeMark and baseline models. Results are presented in a similar format as Table~\ref{tab:id}. We summarize observations related to our research question below.

\begin{table*}[h]
\centering
\renewcommand{\arraystretch}{1.1}
\caption{
Attribution accuracy results on unseen artifacts across distortions and attacks.
}
\resizebox{0.99\textwidth}{!}{%
\begin{tabular}{llc cc cc cc}
\toprule
\multirow{3}{*}{\textbf{}} & 
 \multirow{2}{*}{\diagbox{Distortion}{System}}
& \multicolumn{2}{c}{Proposed Method} 
& \multicolumn{2}{c}{Watermarking Baselines} 
& \multicolumn{2}{c}{Classifier Baselines} \\
\cmidrule(lr){3-4} \cmidrule(lr){5-6} \cmidrule(lr){7-8}
& & \textbf{FakeMark$^{A}$} & \textbf{FakeMark$^{T}$} & \textbf{AudioSeal} & \textbf{Timbre} 
& \textbf{MMS-300M} & \textbf{ResNet34} \\
\midrule
 & None             & \cellcolor[rgb]{1.00, 1.00, 1.00} 1.00 & \cellcolor[rgb]{1.00, 1.00, 1.00} 1.00 & \cellcolor[rgb]{1.00, 1.00, 1.00} 1.00 & \cellcolor[rgb]{1.00, 1.00, 1.00} 1.00 & \cellcolor[rgb]{0.59, 0.59, 0.59} 0.07 & \cellcolor[rgb]{0.62, 0.62, 0.62} 0.12\\
\midrule
\multirow{3}{*}{Signal Processing} 
& Pitch      & \cellcolor[rgb]{0.95, 0.95, 0.95} 0.80 & \cellcolor[rgb]{1.00, 1.00, 1.00} 1.00 & \cellcolor[rgb]{0.93, 0.93, 0.93} 0.72 & \cellcolor[rgb]{0.99, 0.99, 0.99} 0.96 & \cellcolor[rgb]{0.59, 0.59, 0.59} 0.00 & \cellcolor[rgb]{0.65, 0.65, 0.65} 0.10\\ 
& Speed      & \cellcolor[rgb]{1.00, 1.00, 1.00} 0.99 & \cellcolor[rgb]{1.00, 1.00, 1.00} 1.00 & \cellcolor[rgb]{0.95, 0.95, 0.95} 0.78 & \cellcolor[rgb]{1.00, 1.00, 1.00} 0.98 & \cellcolor[rgb]{0.59, 0.59, 0.59} 0.06 & \cellcolor[rgb]{0.62, 0.62, 0.62} 0.11\\
& Noise      & \cellcolor[rgb]{0.97, 0.97, 0.97} 0.58 & \cellcolor[rgb]{0.99, 0.99, 0.99} 0.63 & \cellcolor[rgb]{1.00, 1.00, 1.00} 0.65 & \cellcolor[rgb]{0.99, 0.99, 0.99} 0.62 & \cellcolor[rgb]{0.59, 0.59, 0.59} 0.03 & \cellcolor[rgb]{0.61, 0.61, 0.61} 0.05\\
\midrule
\multirow{3}{*}{Codec} 
& SpeechTokenizer & \cellcolor[rgb]{0.89, 0.89, 0.89} 0.58 & \cellcolor[rgb]{0.99, 0.99, 0.99} 0.88 & \cellcolor[rgb]{0.59, 0.59, 0.59} 0.07 & \cellcolor[rgb]{1.00, 1.00, 1.00} 0.90 & \cellcolor[rgb]{0.59, 0.59, 0.59} 0.07 & \cellcolor[rgb]{0.61, 0.61, 0.61} 0.10\\
& FACodec    & \cellcolor[rgb]{1.00, 1.00, 1.00} 0.87 & \cellcolor[rgb]{1.00, 1.00, 1.00} 0.88 & \cellcolor[rgb]{0.61, 0.61, 0.61} 0.08 & \cellcolor[rgb]{0.99, 0.99, 0.99} 0.85 & \cellcolor[rgb]{0.61, 0.61, 0.61} 0.08 & \cellcolor[rgb]{0.59, 0.59, 0.59} 0.05\\
& WavTokenizer & \cellcolor[rgb]{0.69, 0.69, 0.69} 0.06 & \cellcolor[rgb]{0.83, 0.83, 0.83} 0.11 & \cellcolor[rgb]{0.59, 0.59, 0.59} 0.03 & \cellcolor[rgb]{1.00, 1.00, 1.00} 0.21 & \cellcolor[rgb]{0.72, 0.72, 0.72} 0.07 & \cellcolor[rgb]{0.72, 0.72, 0.72} 0.07\\
\midrule
\multirow{3}{*}{Vocoder} 
& HiFi-GAN   & \cellcolor[rgb]{0.97, 0.97, 0.97} 0.88 & \cellcolor[rgb]{1.00, 1.00, 1.00} 0.98 & \cellcolor[rgb]{0.60, 0.60, 0.60} 0.08 & \cellcolor[rgb]{1.00, 1.00, 1.00} 1.00 & \cellcolor[rgb]{0.59, 0.59, 0.59} 0.07 & \cellcolor[rgb]{0.61, 0.61, 0.61} 0.11\\
& Vocos      & \cellcolor[rgb]{1.00, 1.00, 1.00} 0.98 & \cellcolor[rgb]{1.00, 1.00, 1.00} 1.00 & \cellcolor[rgb]{0.62, 0.62, 0.62} 0.09 & \cellcolor[rgb]{1.00, 1.00, 1.00} 1.00 & \cellcolor[rgb]{0.59, 0.59, 0.59} 0.03 & \cellcolor[rgb]{0.64, 0.64, 0.64} 0.11\\
& BigVGAN    & \cellcolor[rgb]{1.00, 1.00, 1.00} 1.00 & \cellcolor[rgb]{1.00, 1.00, 1.00} 1.00 & \cellcolor[rgb]{0.67, 0.67, 0.67} 0.19 & \cellcolor[rgb]{1.00, 1.00, 1.00} 1.00 & \cellcolor[rgb]{0.59, 0.59, 0.59} 0.06 & \cellcolor[rgb]{0.62, 0.62, 0.62} 0.11\\
\midrule
\multirow{2}{*}{Removal Attack}
& Overwriting & \cellcolor[rgb]{1.00, 1.00, 1.00} 0.97 & \cellcolor[rgb]{0.95, 0.95, 0.95} 0.77 & \cellcolor[rgb]{0.93, 0.93, 0.93} 0.70 & \cellcolor[rgb]{0.87, 0.87, 0.87} 0.54 & \cellcolor[rgb]{0.59, 0.59, 0.59} 0.03 & \cellcolor[rgb]{0.60, 0.60, 0.60} 0.05\\
& Averaging   & \cellcolor[rgb]{1.00, 1.00, 1.00} 0.99 & \cellcolor[rgb]{1.00, 1.00, 1.00} 1.00 & \cellcolor[rgb]{0.93, 0.93, 0.93} 0.73 & \cellcolor[rgb]{1.00, 1.00, 1.00} 1.00 & \cellcolor[rgb]{0.59, 0.59, 0.59} 0.06 & \cellcolor[rgb]{0.61, 0.61, 0.61} 0.10\\
\bottomrule
\end{tabular}
}
\label{tab:od}
\end{table*}

\textbf{FakeMark performs robustly under domain shift.} Under clean conditions, both FakeMark variants and the watermarking baselines achieve perfect accuracy (1.0), this is consistent with their in-domain results in Table~\ref{tab:id}. The two classifier-based models perform poorly (0.07 for MMS and 0.12 for ResNet34), likely due to their limited generalization to unseen artifact patterns—even those produced by TTS architectures seen during training. The two classifiers give similar performance under distortions and attacks, not because their robustness improves in these conditions, but rather because their clean-condition accuracy is already very low.

As in Table~\ref{tab:od}, the performance of FakeMark and watermarking baselines degrades when input signals are distorted, but the trends remain similar to those in Table~\ref{tab:id}, with a slight drop in overall accuracy. Given that MMS-300M fails on this domain-shifted data (highest accuracy 0.08), the robustness of FakeMark detector (above 0.80 under most distortions) can be attributed primarily to the watermarks injected by its generator. Unlike classifier-based models, FakeMark detector is influenced more by distortions applied to the carrier signal than by the carrier itself.

\textbf{Watermarks injected by FakeMark are robust to removal attacks.} From the last two rows of Table~\ref{tab:id}, we may tentatively hypothesize that FakeMark’s robustness against watermark removal attacks was due to the persistence of artifacts. However, results from Table~\ref{tab:od} show, when such artifacts are absent in cross-dataset evaluation, both FakeMark variants remain the most robust among watermarking models (lowest accuracy 0.77 under Overwriting, compared to 0.70 for AudioSeal and 0.54 for Timbre). Hence, this robustness is likely to stem from the injection and detection of watermark message, which is designed to correlate with acoustic artifacts. In contrast, removal attacks primarily focus on removing or overwriting fixed patterns in the carrier signal~\citep{yang2024can}.

\textbf{Additional discussion on watermarking in deepfake attribution.} Both FakeMark variants and Timbre outperform the two classifiers in nearly all test cases across in-domain (Table~\ref{tab:id}) and cross-dataset (Table~\ref{tab:od}) evaluations. Beyond the robustness provided by the system design and training strategies, it is important to note that these solutions are designed for different application scenarios. Classifier-based solutions are passive and require no prior knowledge of the input signal, whereas watermarking-based solutions are proactive and require a message to be injected into the detector input in advance. In the following sections, we further assess the impact of the injected messages on speech quality (Sec.~\ref{sec:subj}) and detector performance (Sec.~\ref{sec:abla}).

\subsubsection{Evaluation on speech quality and intelligibility}
\label{sec:subj}

We evaluate the quality and intelligibility of watermarked signals. Results are presented in Table~\ref{tab:mos}. Our observations are summarized below.

\begin{table}[h!]
\centering
\setlength{\tabcolsep}{7pt}
\renewcommand{\arraystretch}{1.1}
\caption{Comparison of speech quality and intelligibility on watermarked speech signals generated by FakeMark and watermarking-based baselines.}
\label{tab:mos}
\resizebox{0.70\textwidth}{!}{%
\begin{tabular}{llccccc}
\toprule
& System & SI-SNR $\uparrow$ & PESQ $\uparrow$ & ViSQOL $\uparrow$ & PQ $\uparrow$ \\
\midrule

\multirow{2}{*}{Baselines} & AudioSeal & 36.49 & 4.55 & 4.98 & 6.78 \\
& Timbre & 21.79 & 2.97 & 4.20 & 5.67 \\
\midrule
\multirow{2}{*}{Proposed} & FakeMark$^{A}$ & 35.34 & 3.79 & 4.81 & 6.62 \\
& FakeMark$^{T}$ & 14.97 & 2.83 & 4.41 & 6.18 \\
\bottomrule
\end{tabular}
}
\end{table}

\textbf{FakeMark$^{A}$ achieves second in speech quality.} The FakeMark$^{A}$ performs second only to AudioSeal. Its relatively high SI-SNR (35.34 dB) suggests that the injected watermark has low energy compared to the clean carrier. For other speech quality and fidelity metrics, AudioSeal is the only system achieving a PESQ score above 4 (4.55), while FakeMark$^{A}$ is slightly lower in ViSQOL (4.98 vs. 4.81) and PQ (6.78 vs. 6.62). 

\textbf{Trade-off between robustness and speech quality.} We observe that watermarks injected through spectrogram features (Timbre and FakeMark$^{T}$) introduce more distortions to the carrier speech than the approaches that directly process waveforms (AudioSeal and FakeMark$^{A}$). Their worse speech quality contrasts with our observations on attribution performance in Sec.~\ref{sec:in-domain}, and suggests a trade-off between attribution robustness against distortions and speech quality. The consistent near-perfect performance of Timbre and FakeMark$^{T}$ is achieved through stronger, more perceptually noticeable watermarks that can survive multiple distortions (shown in Figures~\ref{fig:timbre-fig1} and~\ref{fig:fakemarkt-fig1} in Appendix~\ref{apd:figures}). In contrast, AudioSeal’s less perceptible watermark introduces minimal distortion to the carrier but is the most vulnerable among the evaluated models. Our proposed FakeMark$^{A}$ provides strong watermark injection while maintaining relatively high speech quality.

\subsubsection{Impact of watermarks on attribution}
\label{sec:abla}

In this section, we examine the extent to which the injected watermarks improve deepfake traceability. We compare FakeMark detector’s performance on non-watermarked, clean signals and randomly watermarked signals with the results from Tables~\ref{tab:id} and~\ref{tab:od}, where watermarks are chosen to always match the ground-truth system label. Results are presented in Table~\ref{tab:abla}.

\begin{table}[h]
\centering
\renewcommand{\arraystretch}{1.0}
\setlength{\tabcolsep}{7pt} 
\caption{Attribution accuracy results of FakeMark detector under different watermarking conditions.}
\label{tab:abla}
\resizebox{0.95\linewidth}{!}{%
\begin{tabular}{rrcccccc}
\toprule
  & Test set & \multicolumn{3}{c}{MLAAD\_v5} & \multicolumn{3}{c}{ASVspoof + TIMIT-TTS} \\
\cmidrule(lr){2-2} \cmidrule(lr){3-5} \cmidrule(lr){6-8}
Generator & \diagbox{Distortion}{Condition} &  No watermark & Random & Matching & No watermark & Random & Matching \\
\midrule
\multirow{2}{*}{FakeMark$^{A}$} & None             & 1.00 & 1.00 & 1.00 & 0.06 & 1.00 & 1.00 \\
&Others averaged  & 0.73 & 0.80 & 0.86 & 0.03 & 0.77 & 0.79 \\
\midrule
\multirow{2}{*}{FakeMark$^{T}$} & None & 0.99 & 1.00 & 1.00 & 0.05 & 1.00 & 1.00\\
& Others averaged & 0.78 & 0.85 & 0.91 & 0.04 & 0.86 & 0.84\\
\bottomrule
\end{tabular}%
}
\end{table}

\textbf{The injected watermarks improve deepfake traceability.} Similar to the classifier baselines in clean conditions, the standalone FakeMark detector achieves near-perfect accuracy (above 0.99) on the MLAAD\_v5 test set, but drops to 0.73 (FakeMark$^{A}$) and 0.78 (FakeMark$^{T}$) under distortions. Adding watermarks improves attribution accuracy for both variants and test conditions, regardless of whether the watermark is randomly assigned or matches the ground-truth label.

Although attribution achieves perfect accuracy with clean signals for both watermark injection conditions, \textbf{injecting watermarks that match the ground-truth label outperforms randomly assigned labels under distortions.} For in-domain MLAAD\_v5 samples, FakeMark$^{A}$ improves from 0.80 to 0.86, and FakeMark$^{T}$ from 0.85 to 0.91. For cross-dataset samples, improvements are small or even absent (FakeMark$^{T}$ from 0.86 to 0.84), which is expected given that the FakeMark detector primarily depends on the watermark messages when artifact patterns are unseen.

\section{Conclusion}

Motivated by the limitations of both classifier- and watermarking-based solutions for deepfake speech attribution, we proposed a novel watermarking framework FakeMark to enhance deepfake traceability. The core novelty of FakeMark is the injection of artifact-correlated watermarks, which allows the detector to leverage both watermark message and deepfake artifacts for attribution. Our results confirm that such design provides improved generalization and robustness across various seen and unseen datasets and under distortions.

\textbf{Limitations} of this work include considering only fully seen architectures during training and evaluation, which constrains the applicability of watermarking-based attribution when scaling to a large number of unseen deepfake systems. We also observe a trade-off between robustness and speech quality--stronger watermarks introduce more distortions to speech signal. Addressing this trade-off could be an important direction for future work.

\subsubsection*{Acknowledgments}
This study is supported by JST AIP Acceleration Research (JPMJCR24U3) and the JSPS grant (25K24398). This study was partially carried out using the TSUBAME4.0 supercomputer at the Institute of Science Tokyo.

\bibliography{iclr2026_conference}
\bibliographystyle{iclr2026_conference}

\appendix
\section{Appendix}

\subsection{List of augmentations during training}
\label{apd:augmentation}

Below is the list of transformations used as augmentation strategies during all system training in our experiments. They are reproduced from the AudioSeal~\citep{sanroman2024proactive} pipeline. During training, each transformation is selected at random with equal probability and the chosen transform is applied to the current mini-batch. They include:
\begin{enumerate}
    \item EnCodec: Inputs are resamples to 24kHz, compressed and reconstructed with EnCodec~\citep{defossez2022encodec} with $nq=16$, and resampled back to 16kHz.
    \item Speed: Playback speed of input signal is changed randomly between 0.9 and 1.1.
    \item Resample: Inputs are resampled to 32kHz and resampled back to 16kHz.
    \item Echo: A delay and less loud copy of the original is added to the input signal. Delay time is randomly sampled between 0.1 and 0.5 seconds, volume of the copied signal is randomly chosen between 0.1 and 0.5.
    \item White noise: Gaussian noise with standard deviation fixed at 0.001 is added to the input signals.
    \item Pink noise: Pink noise with standard deviation fixed at 0.01 is added to the input signals.
    \item Lowpass filtering: A lowpass filter is applied to the input signal with a cutoff frequency at 5kHz.
    \item Highpass filtering: A highpass filter is applied to the input signal with a cutoff frequency at 500Hz.
    \item Bandpass filtering: A bandpass filter is applied to the input signal with a lower cutoff frequency of 300Hz and an upper cutoff frequency of 8kHz.
    \item Smoothing: Inputs are smoothed using a moving average filter with a variable window size between 2 and 10.
    \item Boost: Amplitude of input signal is multiplied by 1.2.
    \item Duck: Amplitude of input signal is multiplied by 0.8.
    \item AAC: Input signal is encoded in AAC format at 128kbps bitrate.
    \item MP3: Input signal is encoded in MP3 format at 128kbps bitrate.
    \item Identity: Returns the unprocessed input signal.
\end{enumerate}

\subsection{FakeMark module architectures}
\label{apd:architecture}
\paragraph{FakeMark$^{A}$}
We adopt the original AudioSeal generator architecture. The encoder uses a 1D convolution (32 channels, kernel size 7) followed by four convolutional blocks, each containing a residual unit (two kernel-3 convolutions with skip connection, doubling channels during down-sampling) and a down-sampling convolution (stride $S$, kernel $K = 2S$; $S = 2,4,5,8$). It concludes with a two-layer LSTM and a final 1D convolution (128 channels, kernel 7) using ELU activations. The decoder mirrors the encoder with transposed convolutions and reversed strides. The latent dimension $H$ is 128.

\paragraph{FakeMark$^{T}$}
We adopt the Timbre encoder architecture but with larger size. A 1024-point Short-Time Fourier Transform (STFT) with 256 hop length is applied to obtain the magnitude spectrogram and phase of the input signal. The magnitude is fed to the 4-layer Carrier Encoder to obtain the encoded carrier feature, which is then concatenated with the original magnitude and the repeated watermark embedding $\mE_{w}$. This combined feature is passed to the 5-layer Watermark Embedder to generate the magnitude spectrogram of watermark signal. The watermarked magnitude spectrogram is obtained by adding watermark magnitude with original clean magnitude. This is different to the original Timber implementation where the Watermark Embedder directly outputs the watermarked magnitude. The watermarked signal is reconstructed via inverse STFT using the original phase and watermarked magnitude. The same original phase is also used for generating watermark waveform with watermark magnitude. The latent dimension $H$ is 513.

\paragraph{Detector}
We use an identical detector architecture for both FakeMark generators. The detector contains a pre-trained wav2vec model (namely the MMS-300M) as front-end. It extract a 1024-dimensional sequence-level representations from the input signal. These representations are then passed through a global average pooling layer to aggregate temporal information, followed by a fully connected layer that produced the output probabilities of 12 classes.

\subsection{Datasets details}
\label{apd:dataset}

Both the MLAAD\_v5 dataset and source tracing challenge protocol can be downloaded from~\url{https://deepfake-total.com/sourcetracing}.

The ASVspoof5 dataset can be downloaded from~\url{https://huggingface.co/datasets/jungjee/asvspoof5}. 

The TIMIT-TTS dataset can be downloaded from~\url{https://zenodo.org/records/6560159}.

\begin{table}[h]
\centering
\caption{Summary of TTS models, Class ID, watermark bits, and number of samples in train, validation, and test sets of MLAAD\_v5 dataset.}
\label{tab:tts_summary_full_wm}
\resizebox{0.99\textwidth}{!}{%
\begin{tabular}{lccccc}
\toprule
\textbf{TTS Model} & \textbf{Class ID} & \textbf{Watermark Bits} & \textbf{Train} & \textbf{Validation} & \textbf{Test} \\
\midrule
Mars5 & 0 & (0,1,0,0) & 275 & 23 & 300 \\
MeloTTS & 1 & (0,0,1,0) & 274 & 22 & 300 \\
Metavoice-1B & 2 & (1,1,1,0) & 267 & 29 & 300 \\
facebook-mms-tts-deu & 3 & (1,1,0,0) & 265 & 31 & 300 \\
tts\_models-en-ljspeech-fast\_pitch & 4 & (1,0,1,1) & 277 & 23 & 0 \\
tts\_models-it-mai\_female-glow-tts & 5 & (1,0,1,0) & 277 & 18 & 0 \\
griffin\_lim & 6 & (0,1,1,1) & 1359 & 125 & 300 \\
suno-bark & 7 & (0,0,0,1) & 137 & 16 & 79 \\
suno-bark-small & 7 & (0,0,0,1) & 126 & 19 & 221 \\
tts\_models-en-ljspeech-tacotron2-DCA & 8 & (1,1,1,1) & 272 & 25 & 49 \\
tts\_models-fr-mai-tacotron2-DDC & 8 & (1,1,1,1) & 264 & 34 & 65 \\
tts\_models-de-thorsten-tacotron2-DDC & 8 & (1,1,1,1) & 261 & 36 & 64 \\
tts\_models-en-ljspeech-tacotron2-DDC & 8 & (1,1,1,1) & 142 & 11 & 32 \\
tts\_models-en-ljspeech-tacotron2-DDC\_ph & 8 & (1,1,1,1) & 135 & 11 & 90 \\
tts\_models-en-ljspeech-speedy-speech & 9 & (1,0,0,0) & 268 & 28 & 0 \\
tts\_models-it-mai\_male-vits & 10 & (0,0,1,1) & 272 & 26 & 44 \\
tts\_models-fr-css10-vits & 10 & (0,0,1,1) & 270 & 27 & 62 \\
tts\_models-it-mai\_female-vits & 10 & (0,0,1,1) & 269 & 29 & 60 \\
tts\_models-lt-cv-vits & 10 & (0,0,1,1) & 264 & 34 & 53 \\
tts\_models-de-css10-vits-neon & 10 & (0,0,1,1) & 264 & 35 & 60 \\
tts\_models-en-ljspeech-vits--neon & 10 & (0,0,1,1) & 261 & 37 & 21 \\
tts\_models-multilingual-multi-dataset-xtts\_v2 & 11 & (1,1,0,1) & 1898 & 185 & 154 \\
tts\_models-multilingual-multi-dataset-xtts\_v1.1 & 11 & (1,1,0,1) & 1623 & 157 & 128 \\
vixTTS & 11 & (1,1,0,1) & 280 & 19 & 18 \\
\bottomrule
\end{tabular}
}
\label{tab:mlaad}
\end{table}

\begin{table}[h]
\centering
\caption{TTS models, source dataset, Class IDs, watermark bits, and sample counts for cross-dataset evaluation.}
\label{tab:tts_samples_with_wm}
\begin{tabular}{lcccc}
\toprule
\textbf{TTS Model} & \textbf{Source Dataset} & \textbf{Class ID} & \textbf{Watermark Bits} & \textbf{Number of Samples} \\
\midrule
A01-GlowTTS & ASVspoof5 & 5 & (1,0,1,0) & 160 \\
A07-FastPitch & ASVspoof5 & 4 & (1,0,1,1) & 160 \\
fastpitch & TIMIT-TTS & 4 & (1,0,1,1) & 160 \\
glowtts & TIMIT-TTS & 5 & (1,0,1,0) & 160 \\
A11-Tacotron2 & ASVspoof5 & 8 & (1,1,1,1) & 160 \\
A29-XTTS & ASVspoof5 & 11 & (1,1,0,1) & 160 \\
A08-VITS & ASVspoof5 & 10 & (0,0,1,1) & 137 \\
vits & TIMIT-TTS & 10 & (0,0,1,1) & 23 \\
\bottomrule
\end{tabular}
\label{tab:cross-datasets}
\end{table}

\subsection{Training and implementation details}
\label{apd:training}

\paragraph{AudioSeal}
We use the official AudioSeal implementation from \url{https://github.com/facebookresearch/audioseal}.

\paragraph{Timbre}
We use the official Timbre implementation from \url{https://github.com/TimbreWatermarking/TimbreWatermarking}.

\paragraph{FakeMark}
For FakeMark training, the learning rate was linearly increased to $1 \times 10^{-4}$ over the first 2,000 mini-batches, and then linearly decayed to 0 at the 50,000th mini-batch, where training stops. All input signals were resampled to 16~kHz if necessary. The waveform amplitude of training samples was randomly adjusted according to the Active Speech Level (ASL) based on ITU-T P.56. Training data were dynamically sampled by grouping files of similar durations and zero-padding them to form mini-batches, with a maximum batch duration of 40~seconds. Files longer than 10~seconds were randomly trimmed to durations between 6 and 10~seconds during training.

Validation was performed every 500 mini-batches, and the best model was selected based on the lowest sum of attribution loss and watermark detection loss. Test samples were neither amplitude-adjusted nor trimmed. 

The balancing weights for training were set as follows: attribution loss, 10.0; watermark detection loss, 10.0; HiFi-GAN losses, 1.0 (with $L_1$ spectrogram loss weight 1.0 and feature matching loss weight 1.0); AudioSeal perceptual losses, 0.1 for $L_1$ loss, 10.0 for loudness loss, and 1.0 for frequency magnitude loss.

AudioSeal was trained on 6 NVIDIA A100 GPUs. The left training were performed on a single NVIDIA H100 GPU.

\paragraph{MMS-300M Classifier}
We adopt the same architecture as the FakeMark detector and use the same codebase and training procedure, except that the maximum learning rate is set to $1 \times 10^{-5}$ and the batch size is fixed at 16. Training stops after 30,000 mini-batches. Best model selection is based on the classification accuracy on validation set.

\paragraph{ResNet34 Classifier}
We use a standard ResNet34 architecture with a temporal statistics pooling layer (TSPL) to extract a 128-dimensional embedding from the input signal, followed by a fully connected layer for prediction. The input is a randomly selected 4-second segment of the original signal, padded if shorter. Following~\citet{klein2025open}, we use 80-dimensional log linear filter-bank (LFB) features of the speech signal, computed with a 400-sample window, 160-sample hop, and a 400-point FFT. We further compute delta ($\Delta$) and double-delta ($\Delta\Delta$) features, and apply cepstral mean and variance normalization (CMVN), yielding a final feature dimension of 240. The model is trained using the Large Margin Cosine Loss with default settings from the implementation in~\url{https://github.com/YirongMao/softmax_variants/blob/master/model_utils.py#L103}. All hyperparameters are identical to those used for MMS-300M training, except that the maximum learning rate is set to $1 \times 10^{-4}$.

\subsection{List of distortions during evaluation}
\label{apd:distortion}
The settings of distortion and watermark removal attacks are:
\begin{enumerate}
    \item Pitch shift: Pitch is randomly shifted  between -1 and 1 semitones.
    \item Playback speed: Original speed is adjust to a number randomly sampled between 0.95 and 1.05.
    \item Noise: Random noise from MUSAN noise recordings is applied at 0dB SNR.
    \item BigVGAN: Using code and pre-trained weight from \url{https://github.com/NVIDIA/BigVGAN}. Input signals are resampled to 24kHz, passed to BigVGAN vocoder, and resampled back to 16kHz.
    \item HiFi-GAN: Using the pre-trained weights from \url{https://huggingface.co/speechbrain/tts-hifigan-libritts-16kHz}.
    \item Vocos: Using code and pre-trained weight (vocos-mel-24khz) from \url{https://github.com/gemelo-ai/vocos/tree/main}. Input signals are resampled to 24kHz, passed to Vocos, and resampled back to 16kHz.
    \item SpeechTokenizer: Using code and pre-trained weight (speechtokenizer\_hubert\_avg) from \url{https://github.com/ZhangXInFD/SpeechTokenizer}.
    \item FACodec: Using code and pre-trained weight from \url{https://huggingface.co/amphion/naturalspeech3_facodec}.
    \item WavTokenizer: Using code and pre-trained weight (WavTokenizer-small-600-24k-4096) from \url{https://huggingface.co/amphion/naturalspeech3_facodec}. Input signals are resampled to 24kHz, passed to WavTokenizer, and resampled back to 16kHz.
    \item Overwriting: Input signals are sequentially passed through pre-trained Timbre and AudioSeal models three times to obtain the watermarked signal.
    \item Averaging: Data samples from the zh-CN subset of the Common Voice dataset are processed using the pre-trained AudioSeal model. The resulting watermark signals for each sample are summed and averaged, and this averaged watermark is then subtracted from the input signal. We did not apply the Averaging attack with pre-trained Timbre model because its generator directly outputs the watermarked signal rather than estimating a separate watermark.
\end{enumerate}

\subsection{Visualizations of speech signals}
\label{apd:figures}

\begin{figure}
    \centering
    \includegraphics[width=1\linewidth]{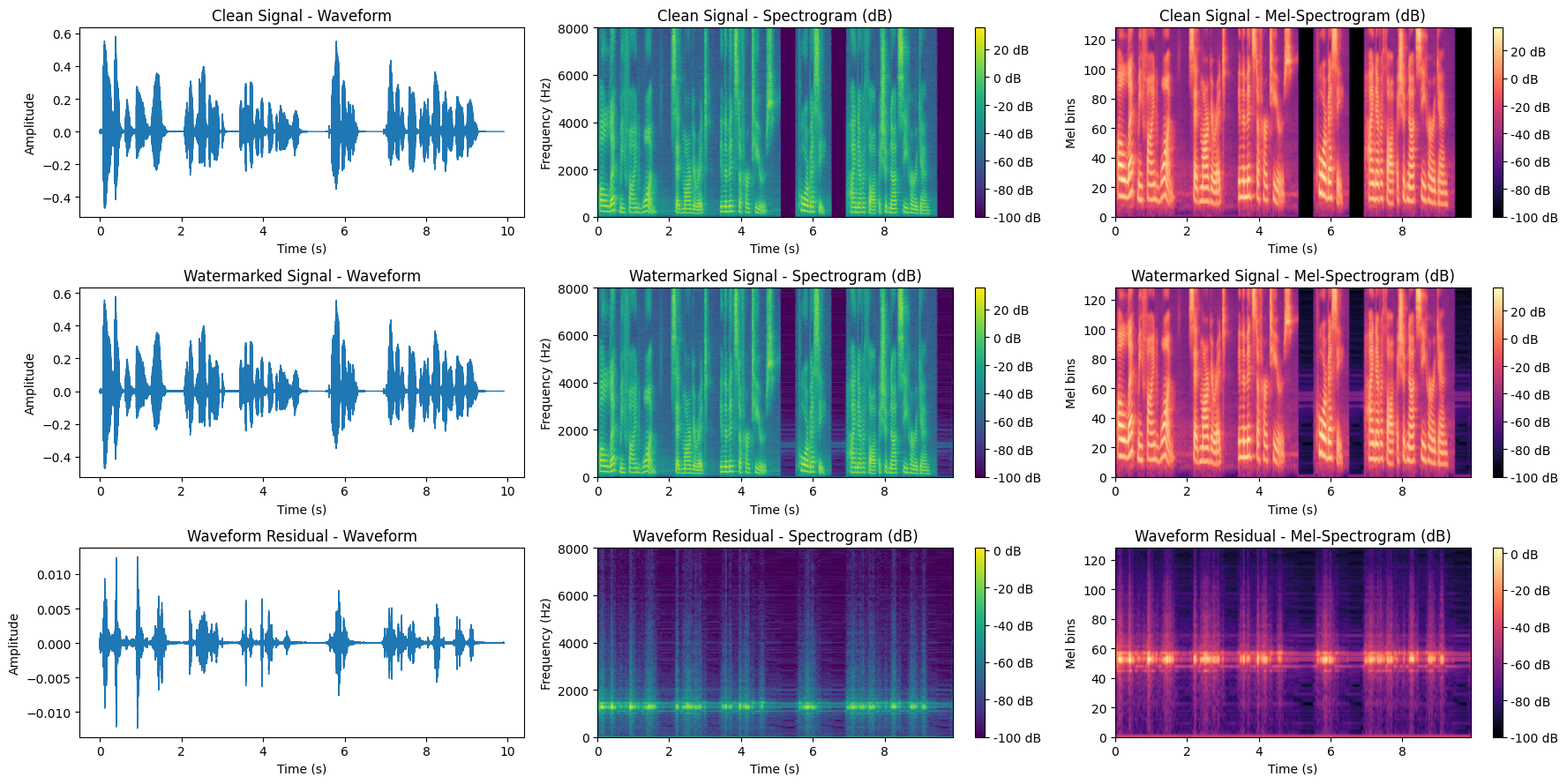}
    \caption{Visualization of AudioSeal watermarking on MLAAD-en-tts\_models-en-ljspeech-tacotron2-DDC-northandsouth\_27\_f000104.}
    \label{fig:audioseal-fig1}
\end{figure}

\begin{figure}
    \centering
    \includegraphics[width=1\linewidth]{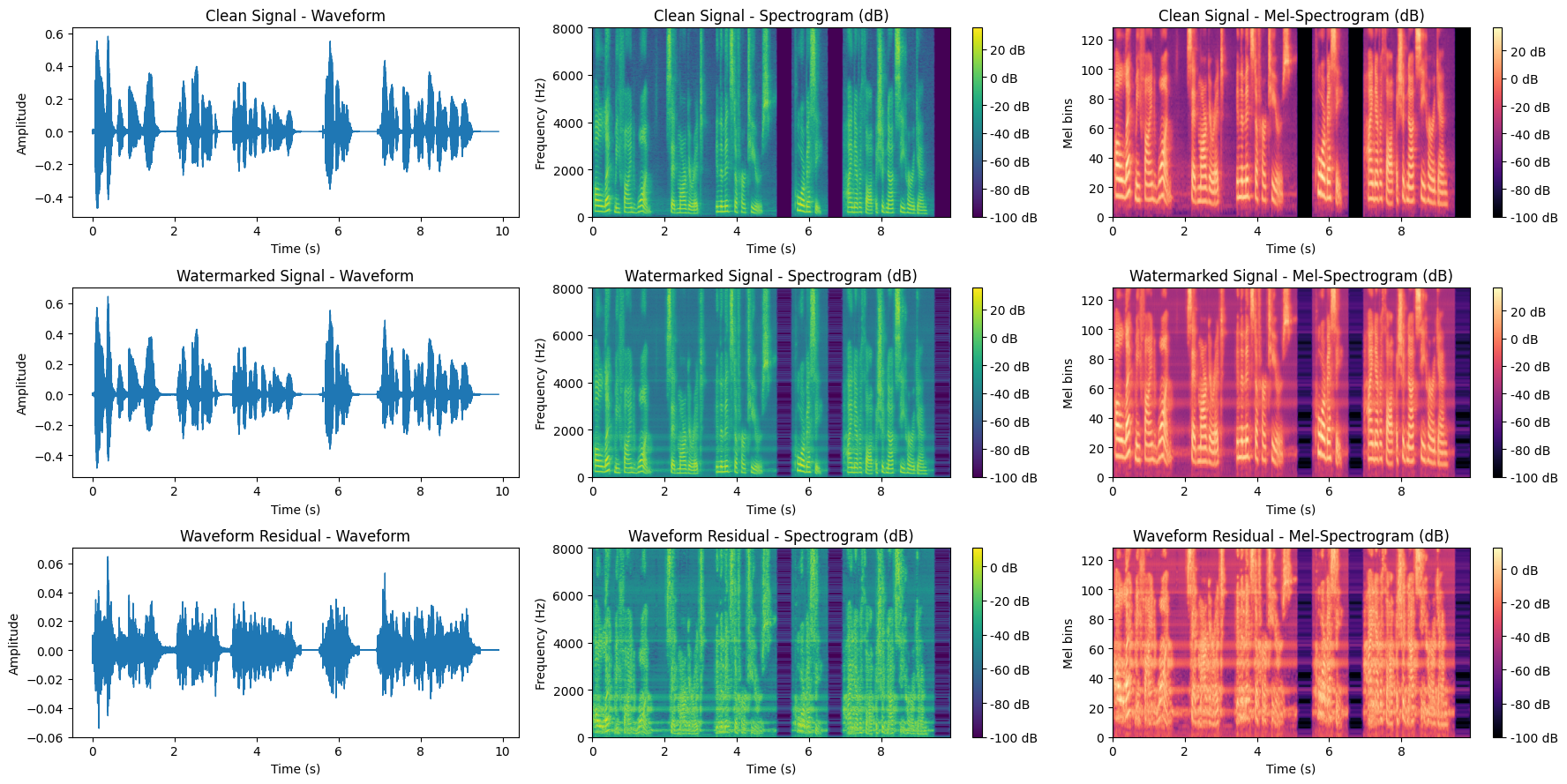}
    \caption{Visualization of Timbre watermarking on MLAAD-en-tts\_models-en-ljspeech-tacotron2-DDC-northandsouth\_27\_f000104.}
    \label{fig:timbre-fig1}
\end{figure}

\begin{figure}
    \centering
    \includegraphics[width=1\linewidth]{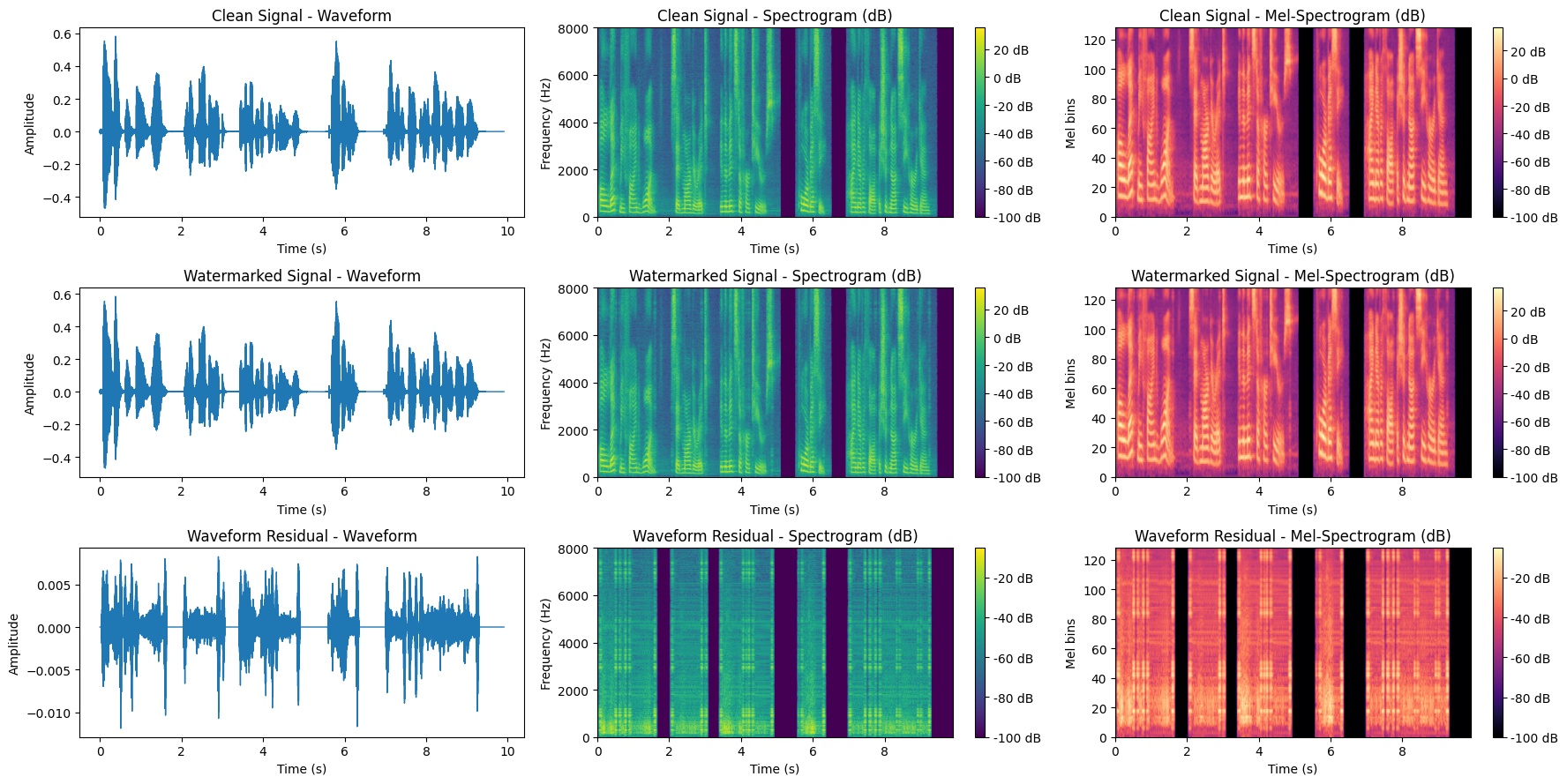}
    \caption{Visualization of FakeMark$^{A}$ watermarking on MLAAD-en-tts\_models-en-ljspeech-tacotron2-DDC-northandsouth\_27\_f000104.}
    \label{fig:fakemarka-fig1}
\end{figure}

\begin{figure}
    \centering
    \includegraphics[width=1\linewidth]{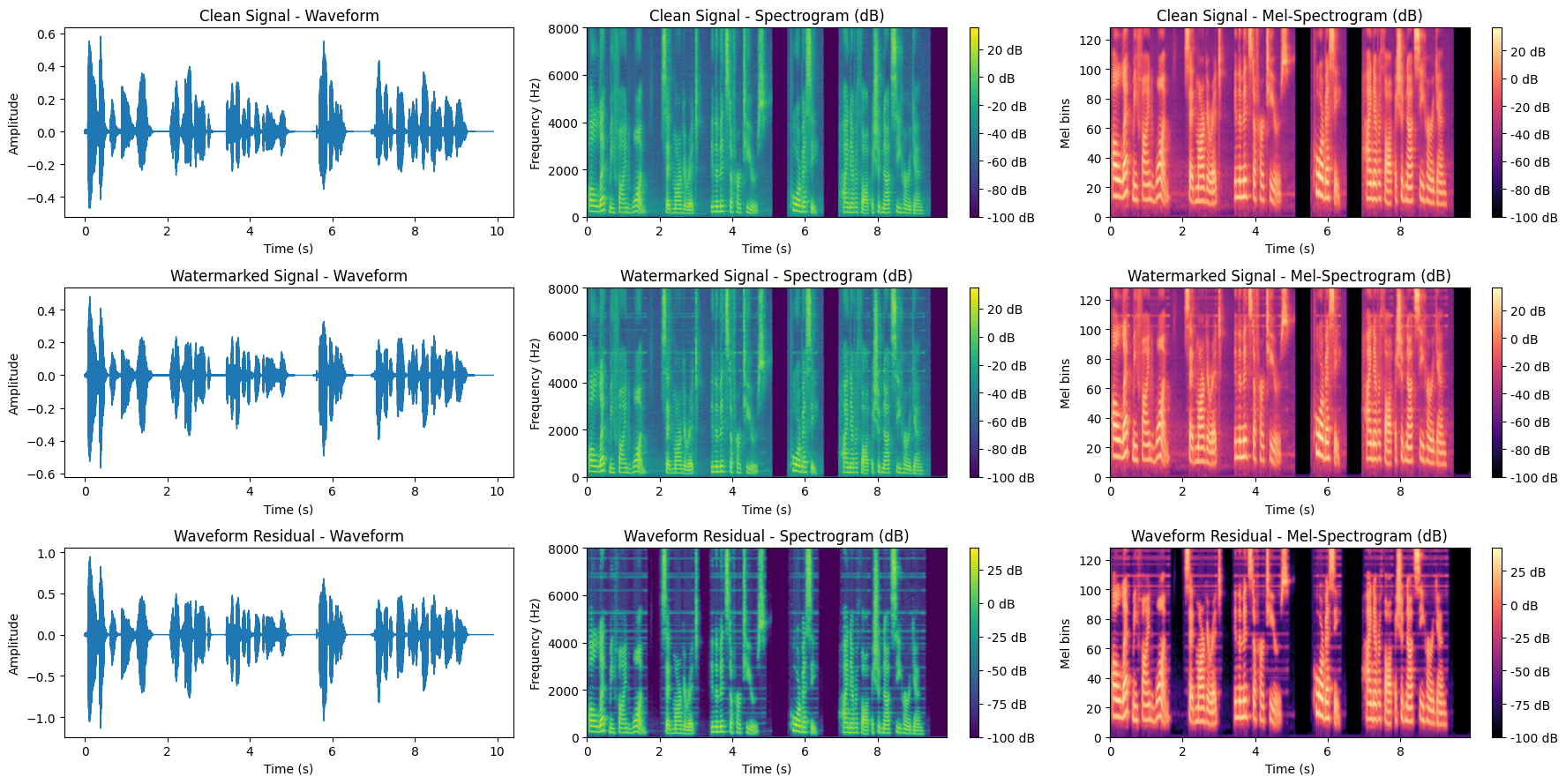}
    \caption{Visualization of FakeMark$^{T}$ watermarking on MLAAD-en-tts\_models-en-ljspeech-tacotron2-DDC-northandsouth\_27\_f000104.}
    \label{fig:fakemarkt-fig1}
\end{figure}

\begin{figure}
    \centering
    \includegraphics[width=1\linewidth]{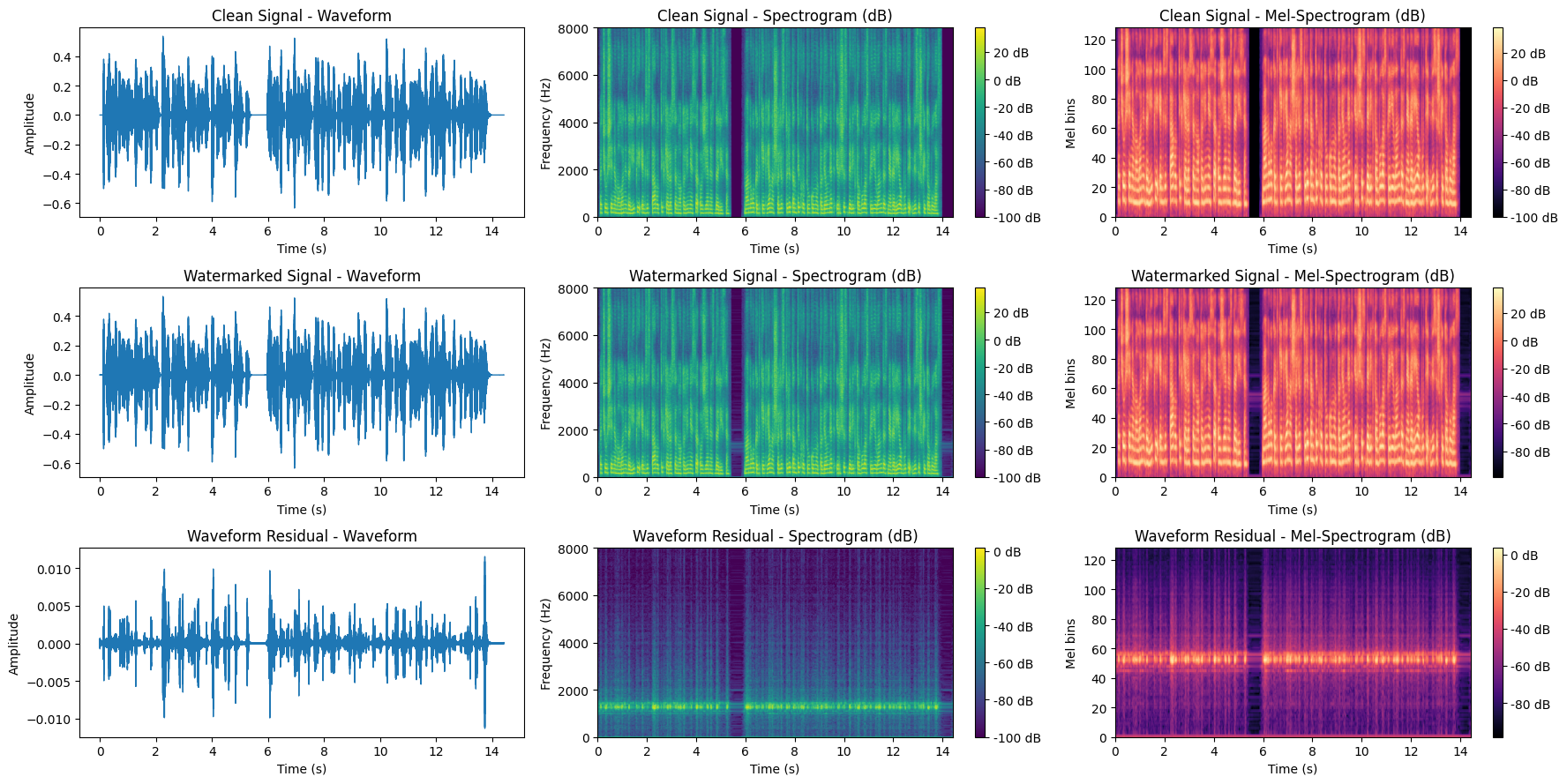}
    \caption{Visualization of AudioSeal watermarking on MLAAD-lt-tts\_models-lt-cv-vits-emerald\_city\_of\_oz\_03\_f000037.}
    \label{fig:audioseal-fig2}
\end{figure}

\begin{figure}
    \centering
    \includegraphics[width=1\linewidth]{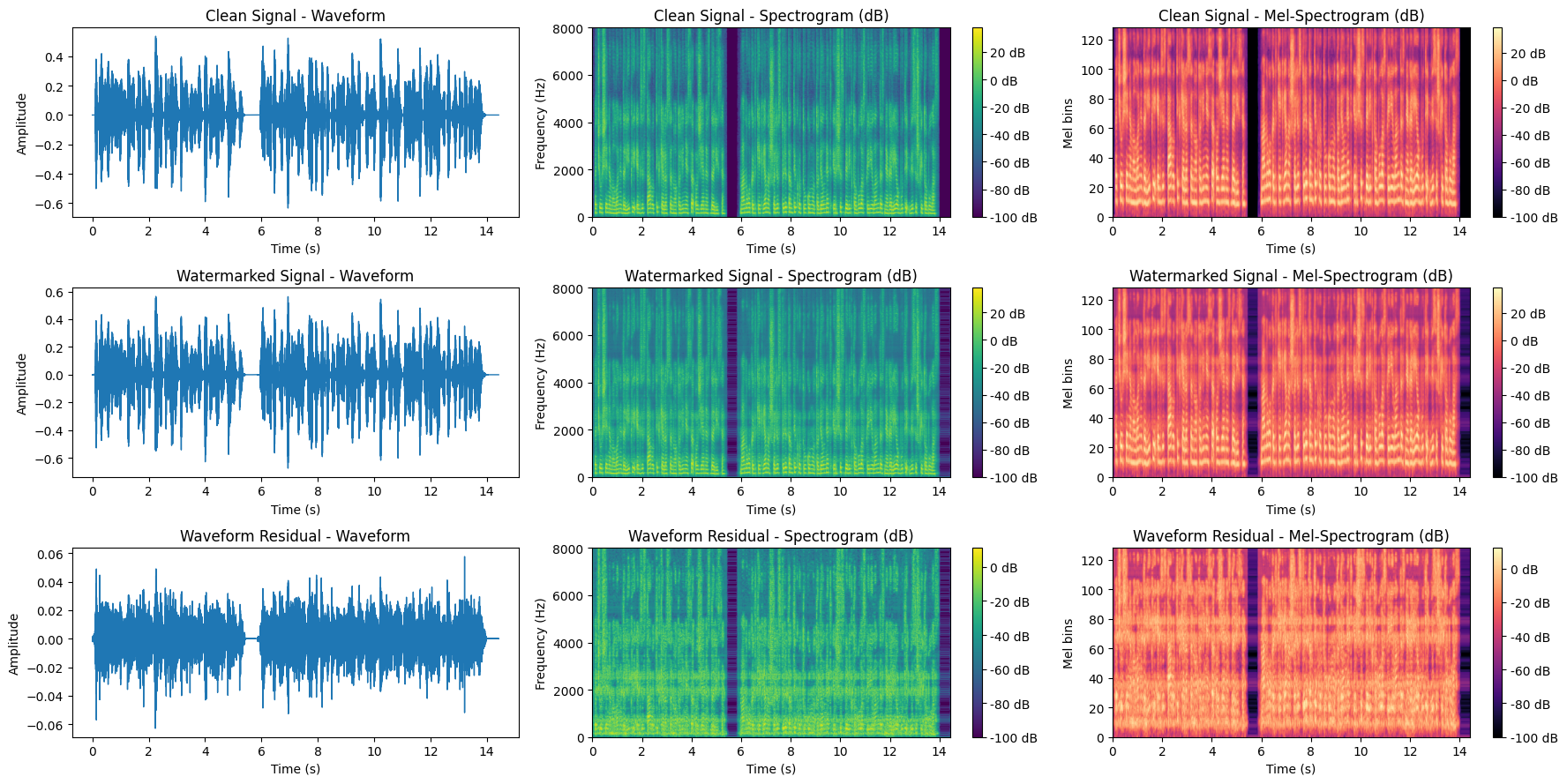}
    \caption{Visualization of Timbre watermarking on MLAAD-lt-tts\_models-lt-cv-vits-emerald\_city\_of\_oz\_03\_f000037.}
    \label{fig:timbre-fig2}
\end{figure}

\begin{figure}
    \centering
    \includegraphics[width=1\linewidth]{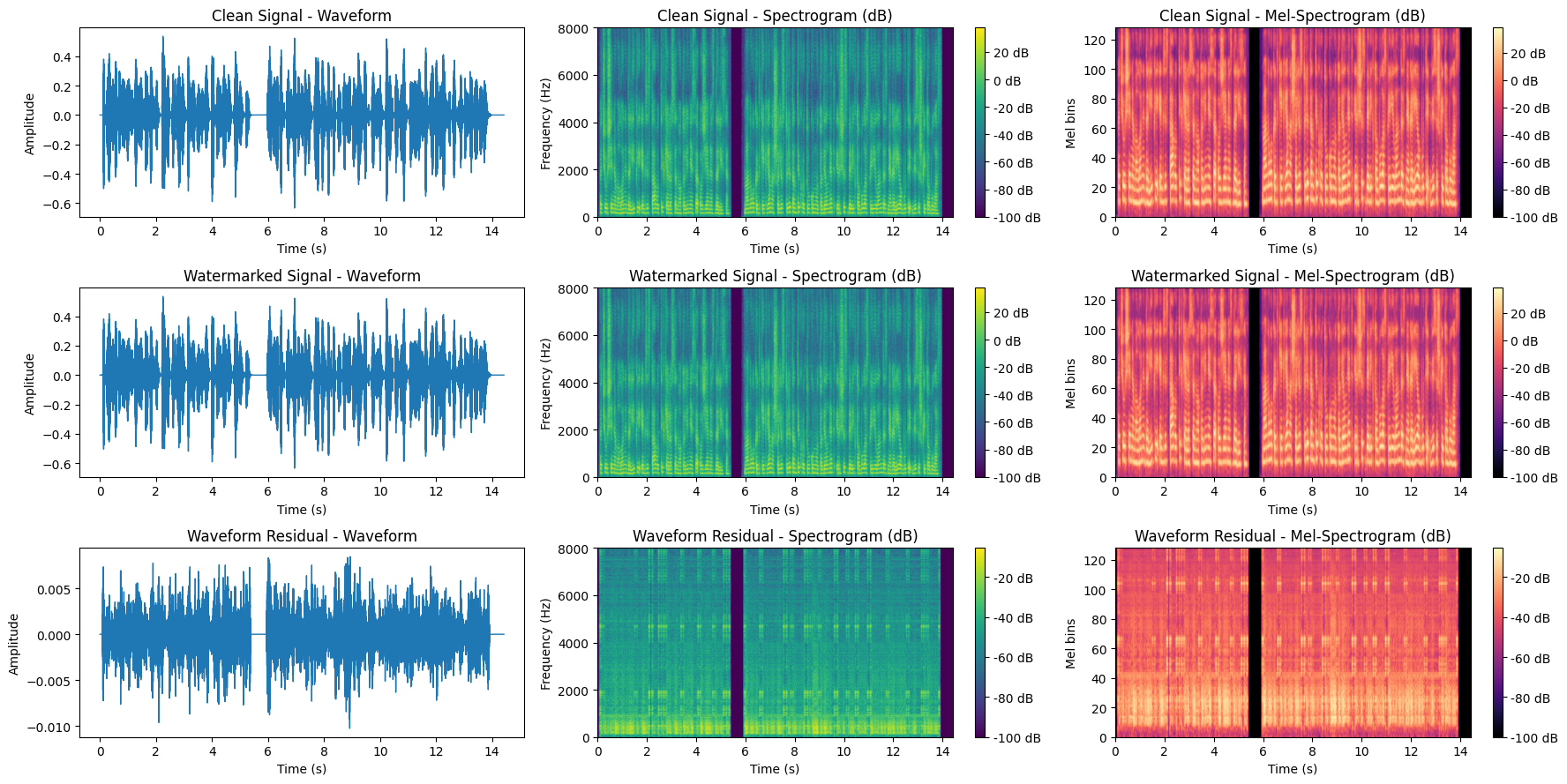}
    \caption{Visualization of FakeMark$^{A}$ watermarking on MLAAD-lt-tts\_models-lt-cv-vits-emerald\_city\_of\_oz\_03\_f000037.}
    \label{fig:fakemarka-fig2}
\end{figure}

\begin{figure}
    \centering
    \includegraphics[width=1\linewidth]{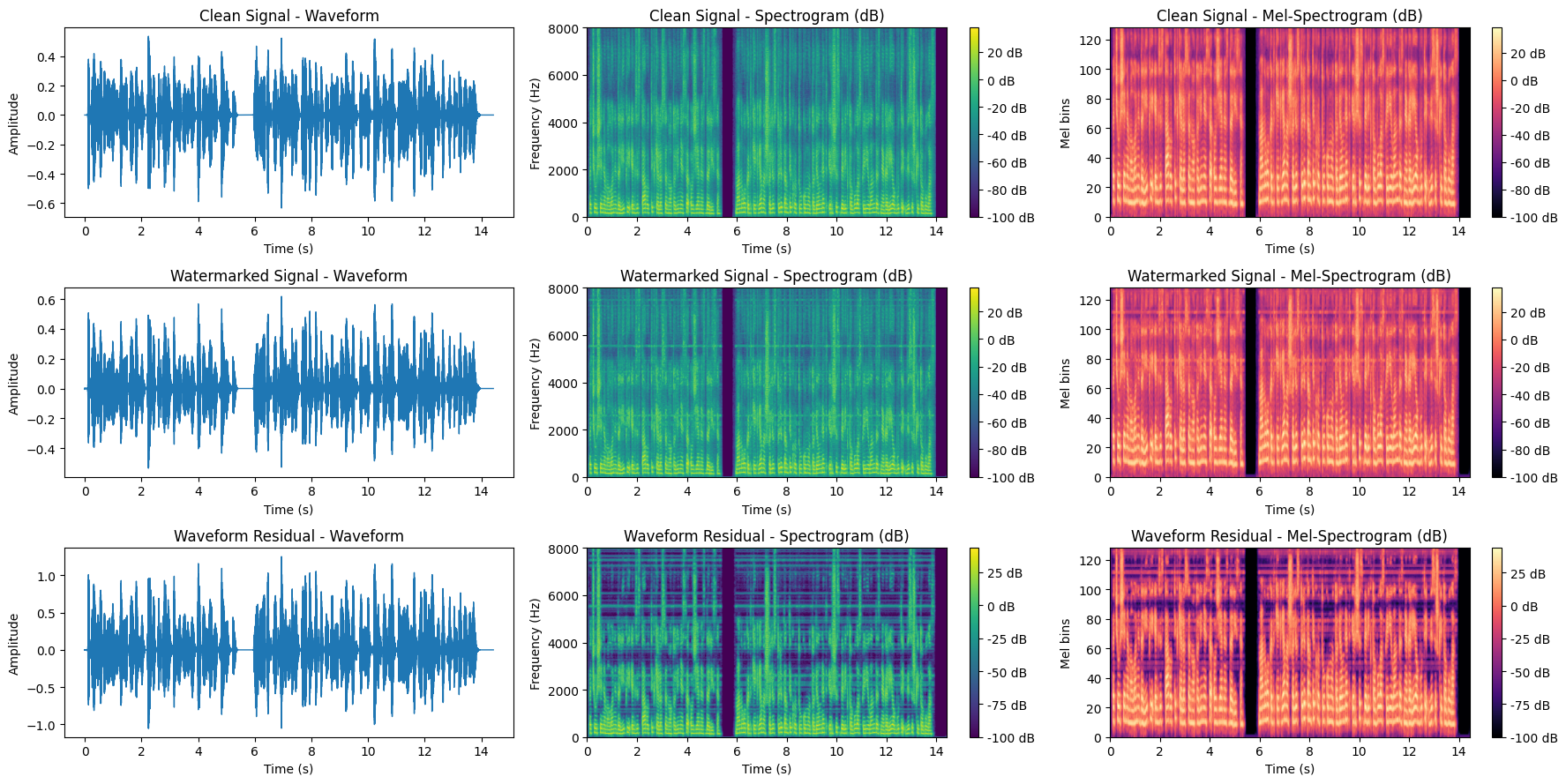}
    \caption{Visualization of FakeMark$^{T}$ watermarking on MLAAD-lt-tts\_models-lt-cv-vits-emerald\_city\_of\_oz\_03\_f000037.}
    \label{fig:fakemarkt-fig2}
\end{figure}

\end{document}